

\def\pmb#1{\setbox0=\hbox{#1}%
  \hbox{\kern-.025em\copy0\kern-\wd0
  \kern.05em\copy0\kern-\wd0
  \kern-0.025em\raise.0433em\box0} }

\catcode`@=11
\def\leftrightarrowfill{$\m@th\mathord\leftarrow \mkern-6mu
  \cleaders\hbox{$\mkern-2mu \mathord- \mkern-2mu$}\hfill
  \mkern-6mu \mathord\rightarrow$}
\def\overleftrightarrow#1{\vbox{\ialign{##\crcr
     \leftrightarrowfill\crcr\noalign{\kern-1pt\nointerlineskip}
     $\hfil\displaystyle{#1}\hfil$\crcr}}}
\catcode`@=12

\def\approxge{\hbox {\hfil\raise .4ex\hbox{$>$}\kern-.75 em
\lower .7ex\hbox{$\sim$}\hfil}}
\def\approxle{\hbox {\hfil\raise .4ex\hbox{$<$}\kern-.75 em
\lower .7ex\hbox{$\sim$}\hfil}}

\def \abstract#1 {\vskip 0.5truecm\sepline\vskip 0.5truecm
$$\vbox{\hsize=15truecm\noindent #1}$$}
\def \SISSA#1#2 {\vfil\vfil\centerline{Ref. S.I.S.S.A. #1 CM (#2)}}
\def \PACS#1 {\vfil\line{\hfil\hbox to 15truecm{PACS numbers: #1 \hfil}\hfil}}

\def \hfigure
     #1#2#3       {\midinsert \vskip #2 truecm $$\vbox{\hsize=14.5truecm
             \seven\baselineskip=10pt\noindent {\bcp \noindent Figure  #1}.
                   #3 } $$ \vskip -20pt \endinsert }

\def \hfiglin
     #1#2#3       {\midinsert \vskip #2 truecm $$\vbox{\hsize=14.5truecm
              \seven\baselineskip=10pt\noindent {\bcp \hfil\noindent
                   Figure  #1}. #3 \hfil} $$ \vskip -20pt \endinsert }

\def \vfigure
     #1#2#3#4     {\dimen0=\hsize \advance\dimen0 by -#3truecm
                   \midinsert \vbox to #2truecm{ \seven
                   \parshape=1 #3truecm \dimen0 \baselineskip=10pt \vfill
                   \noindent{\bcp Figure #1} \pretolerance=6500#4 \vfill }
                   \endinsert }

%
\def \ref
     #1#2         {\smallskip \item{[#1]}#2}
\def \sepline     {\medskip\centerline{\vbox{\hrule width5truecm}} \medskip}

\def \tabrul2     {\noalign{\vskip 5truept \hrule \vskip 2truept \hrule
                   \vskip 5truept} }


\footline={\ifnum\pageno>0 \tenrm \hss \folio \hss \fi }

\def\today
 {\count10=\year\advance\count10 by -1900 \number\day--\ifcase
  \month \or Jan\or Feb\or Mar\or Apr\or May\or Jun\or
             Jul\or Aug\or Sep\or Oct\or Nov\or Dec\fi--\number\count10}

\def\hour{\count10=\time\count11=\count10
\divide\count10 by 60 \count12=\count10
\multiply\count12 by 60 \advance\count11 by -\count12\count12=0
\number\count10 :\ifnum\count11 < 10 \number\count12\fi\number\count11}

\def\draft{
   \baselineskip=20pt
   \def\makeheadline{\vbox to 10pt{\vskip-22.5pt
   \line{\vbox to 8.5pt{}\the\headline}\vss}\nointerlineskip}
   \headline={\hfill \seven {\bcp Draft version}: today is \today\ at \hour
              \hfill}
          }

%
%

%
\catcode`@=11
%
%
\def\b@lank{ }

\newif\if@simboli
\newif\if@riferimenti

\newwrite\file@simboli
\def\simboli{
    \immediate\write16{ !!! Genera il file \jobname.SMB }
    \@simbolitrue\immediate\openout\file@simboli=\jobname.smb}

\newwrite\file@ausiliario
\def\riferimentifuturi{
    \immediate\write16{ !!! Genera il file \jobname.AUX }
    \@riferimentitrue\openin1 \jobname.aux
    \ifeof1\relax\else\closein1\relax\input\jobname.aux\fi
    \immediate\openout\file@ausiliario=\jobname.aux}

\newcount\eq@num\global\eq@num=0
\newcount\sect@num\global\sect@num=0

\newif\if@ndoppia
\def\numerazionedoppia{\@ndoppiatrue\gdef\la@sezionecorrente{\the\sect@num}}

\def\se@indefinito#1{\expandafter\ifx\csname#1\endcsname\relax}
\def\spo@glia#1>{} 

\newif\if@primasezione
\@primasezionetrue

\def\s@ection#1\par{\immediate
    \write16{#1}\if@primasezione\global\@primasezionefalse\else\goodbreak
    \vskip\spaziosoprasez\fi\noindent
    {\bf#1}\nobreak\vskip\spaziosottosez\nobreak\noindent}
%

\def\sezpreset#1{\global\sect@num=#1
    \immediate\write16{ !!! sez-preset = #1 }   }

\def\spaziosoprasez{50pt plus 60pt}
\def\spaziosottosez{15pt}

\def\sref#1{\se@indefinito{@s@#1}\immediate\write16{ ??? \string\sref{#1}
    non definita !!!}
    \expandafter\xdef\csname @s@#1\endcsname{??}\fi\csname @s@#1\endcsname}

\def\autosez#1#2\par{
    \global\advance\sect@num by 1\if@ndoppia\global\eq@num=0\fi
    \xdef\la@sezionecorrente{\the\sect@num}
    \def\usa@getta{1}\se@indefinito{@s@#1}\def\usa@getta{2}\fi
    \expandafter\ifx\csname @s@#1\endcsname\la@sezionecorrente\def
    \usa@getta{2}\fi
    \ifodd\usa@getta\immediate\write16
      { ??? possibili riferimenti errati a \string\sref{#1} !!!}\fi
    \expandafter\xdef\csname @s@#1\endcsname{\la@sezionecorrente}
    \immediate\write16{\la@sezionecorrente. #2}
    \if@simboli
      \immediate\write\file@simboli{ }\immediate\write\file@simboli{ }
      \immediate\write\file@simboli{  Sezione
                                  \la@sezionecorrente :   sref.   #1}
      \immediate\write\file@simboli{ } \fi
    \if@riferimenti
      \immediate\write\file@ausiliario{\string\expandafter\string\edef
      \string\csname\b@lank @s@#1\string\endcsname{\la@sezionecorrente}}\fi
    \goodbreak\vskip 48pt plus 60pt
\centerline{\lltitle #2}                     
\par\nobreak\vskip 15pt \nobreak\noindent}

\def\semiautosez#1#2\par{
    \gdef\la@sezionecorrente{#1}\if@ndoppia\global\eq@num=0\fi
    \if@simboli
      \immediate\write\file@simboli{ }\immediate\write\file@simboli{ }
      \immediate\write\file@simboli{  Sezione ** : sref.
          \expandafter\spo@glia\meaning\la@sezionecorrente}
      \immediate\write\file@simboli{ }\fi
\noindent\lltitle \s@ection#2 \par}


\def\eqpreset#1{\global\eq@num=#1
     \immediate\write16{ !!! eq-preset = #1 }     }

\def\eqref#1{\se@indefinito{@eq@#1}
    \immediate\write16{ ??? \string\eqref{#1} non definita !!!}
    \expandafter\xdef\csname @eq@#1\endcsname{??}
    \fi\csname @eq@#1\endcsname}

\def\eqlabel#1{\global\advance\eq@num by 1
    \if@ndoppia\xdef\il@numero{\la@sezionecorrente.\the\eq@num}
       \else\xdef\il@numero{\the\eq@num}\fi
    \def\usa@getta{1}\se@indefinito{@eq@#1}\def\usa@getta{2}\fi
    \expandafter\ifx\csname @eq@#1\endcsname\il@numero\def\usa@getta{2}\fi
    \ifodd\usa@getta\immediate\write16
       { ??? possibili riferimenti errati a \string\eqref{#1} !!!}\fi
    \expandafter\xdef\csname @eq@#1\endcsname{\il@numero}
    \if@ndoppia
       \def\usa@getta{\expandafter\spo@glia\meaning
       \la@sezionecorrente.\the\eq@num}
       \else\def\usa@getta{\the\eq@num}\fi
    \if@simboli
       \immediate\write\file@simboli{  Equazione
            \usa@getta :  eqref.   #1}\fi
    \if@riferimenti
       \immediate\write\file@ausiliario{\string\expandafter\string\edef
       \string\csname\b@lank @eq@#1\string\endcsname{\usa@getta}}\fi}

\def\autoeqno#1{\eqlabel{#1}\eqno(\csname @eq@#1\endcsname)}
\def\autoleqno#1{\eqlabel{#1}\leqno(\csname @eq@#1\endcsname)}
\def\eqrefp#1{(\eqref{#1})}


\def\eq{\autoeqno}
\def\req{\eqrefp}
\def\chap{\autosez}        



\newcount\cit@num\global\cit@num=0

\newwrite\file@bibliografia
\newif\if@bibliografia
\@bibliografiafalse

\def\lp@cite{[}
\def\rp@cite{]}
\def\trap@cite#1{\lp@cite #1\rp@cite}
\def\lp@bibl{[}
\def\rp@bibl{]}
\def\trap@bibl#1{\lp@bibl #1\rp@bibl}

\def\refe@renza#1{\if@bibliografia\immediate        
    \write\file@bibliografia{
    \string\item{\trap@bibl{\cref{#1}}}\string
    \bibl@ref{#1}\string\bibl@skip}\fi}

\def\ref@ridefinita#1{\if@bibliografia\immediate\write\file@bibliografia{
    \string\item{?? \trap@bibl{\cref{#1}}} ??? tentativo di ridefinire la
      citazione #1 !!! \string\bibl@skip}\fi}

\def\bibl@ref#1{\se@indefinito{@ref@#1}\immediate
    \write16{ ??? biblitem #1 indefinito !!!}\expandafter\xdef
    \csname @ref@#1\endcsname{ ??}\fi\csname @ref@#1\endcsname}

\def\c@label#1{\global\advance\cit@num by 1\xdef            
   \la@citazione{\the\cit@num}\expandafter
   \xdef\csname @c@#1\endcsname{\la@citazione}}

\def\bibl@skip{\vskip +4truept}


\def\stileincite#1#2{\global\def\lp@cite{#1}\global   
    \def\rp@cite{#2}}                                 
\def\stileinbibl#1#2{\global\def\lp@bibl{#1}\global   
    \def\rp@bibl{#2}}                                 

\def\citpreset#1{\global\cit@num=#1
    \immediate\write16{ !!! cit-preset = #1 }    }

\def\autobibliografia{\global\@bibliografiatrue\immediate
    \write16{ !!! Genera il file \jobname.BIB}\immediate
    \openout\file@bibliografia=\jobname.bib}

\def\cref#1{\se@indefinito                  
   {@c@#1}\c@label{#1}\refe@renza{#1}\fi\csname @c@#1\endcsname}

\def\cite#1{\trap@cite{\cref{#1}}}                  
\def\ccite#1#2{\trap@cite{\cref{#1},\cref{#2}}}     
\def\ncite#1#2{\trap@cite{\cref{#1}--\cref{#2}}}    
\def\upcite#1{$^{\,\trap@cite{\cref{#1}}}$}               
\def\upccite#1#2{$^{\,\trap@cite{\cref{#1},\cref{#2}}}$}  
\def\upncite#1#2{$^{\,\trap@cite{\cref{#1}-\cref{#2}}}$}  

\def\clabel#1{\se@indefinito{@c@#1}\c@label           
    {#1}\refe@renza{#1}\else\c@label{#1}\ref@ridefinita{#1}\fi}

\def\biblskip#1{\def\bibl@skip{\vskip #1}}           

\def\insertbibliografia{\if@bibliografia             
    \immediate\write\file@bibliografia{ }
    \immediate\closeout\file@bibliografia
    \catcode`@=11\input\jobname.bib\catcode`@=12\fi}


\def\commento#1{\relax}
\def\biblitem#1#2\par{\expandafter\xdef\csname @ref@#1\endcsname{#2}}


\catcode`@=12


\magnification=1200
\topskip 20pt
\def\interlinea{\baselineskip=16pt}
\def\standardpage{\vsize=20.7truecm\voffset=+1.truecm
                  \hsize=15.truecm\hoffset=+10truemm
                  \parindent=1.2truecm}

\tolerance 100000
\biblskip{+8truept}                        
\def\hbup{\hfill\break\baselineskip 16pt}  


\global\newcount\notenumber \global\notenumber=0
\def\note #1 {\global\advance\notenumber by1 \baselineskip 10pt
              \footnote{$^{\the\notenumber}$}{\nine #1} \interlinea}



\font\text=cmr10
\font\scal=cmsy5
\font\it=cmti10

\font\title=cmbx10 scaled \magstep3      
\font\ltitle=cmbx12 scaled \magstep1
\font\lltitle=cmbx12 scaled \magstep1

\font\abs=cmti10 scaled \magstep1        

\font\seven=cmr7                         
\font\nine=cmr9                         
\font\bcp=cmbx7








\def\gtrsim{\ \rlap{\raise 2pt \hbox{$>$}}{\lower 2pt \hbox{$\sim$}}\ }
\def\lesssim{\ \rlap{\raise 2pt \hbox{$<$}}{\lower 2pt \hbox{$\sim$}}\ }


\def\mn{\medskip\noindent}
\def\bs{\bigskip}
\def\hb{\hfil\break}

\def\o{\over}



\def\ea{{\it et.al.}}
\def\ib{{\it ibid.\ }}

\def\npb#1{Nucl. Phys. {\bf B#1},}
\def\plb#1{Phys. Lett. {\bf B#1},}
\def\prd#1{Phys. Rev. {\bf D#1},}
\def\prl#1{Phys. Rev. Lett. {\bf #1},}

\def\zpc#1{Z. Phys. {\bf C#1},}
\def\prep#1{Phys. Rep. {\bf #1},}


\stileincite{}{}     
\numerazionedoppia   

\interlinea
\standardpage
\text                


\def\o{\over}



\def\G{{\cal G_{\rm SM}}}
\def\E{{\rm E}_6}

\def\N{{\hbox{\scal N}}}
\def\K{{\hbox{\scal K}}}


\def\pr{\prime}


\font\mbf=cmmib10  scaled \magstep1      

\def\bfmu{{\hbox{\mbf\char'026}}}

\def\bfe{{\hbox{\mbf\char'145}}}



\autobibliografia
\def\pha{\phantom{a}}
\pageno=0
\vsize=23.8truecm
\hsize=15.7truecm
\voffset=-1.truecm
\hoffset=+6truemm
\baselineskip 14pt
\rightline{UM-TH 93--08}\par\noindent
\rightline{FTUV 93--14}\par\noindent

\bs\bs\bs
\centerline{\ltitle  \bfmu{} - \bfe{}  conversion in nuclei and
Z$^{\displaystyle\bf \prime}$ physics}
\medskip
\medskip
\bs\bs
\centerline{J. Bernab\'eu$^a$, E. Nardi$^b$ and D. Tommasini$^a$}
\bs
\centerline{\it \pha$^a$ Instituto de F\'\i sica Corpuscular - C.S.I.C.,
and Departament de F\'\i sica Te\` orica}
\centerline{\it Universitat de Val\` encia, 46100 Burjassot, Val\` encia,
SPAIN}
\bs
\centerline{\it \pha$^b$ Randall Laboratory of Physics, University of Michigan,
           Ann Arbor, MI 48109--1120}
\vskip 1truecm                             

\medskip
\centerline  { \bf {\abs Abstract}}      
\bs

\noindent
Together with the existence of new neutral gauge bosons,
models based on extended gauge groups (rank $> 4$) often predict
also new charged fermions.
A mixing of the known fermions with new states with {\it exotic}
weak-isospin assignments (left-handed singlets and right-handed
doublets) will induce tree level flavour changing neutral interactions
mediated by $Z$ exchange, while if the mixing is
only with new states with {\it ordinary} weak-isospin assignments,
the flavour changing neutral currents are mainly due to the exchange
of the lightest new neutral gauge boson $Z^\prime$.
We show that the present experimental limits on $\mu-e$
conversion in nuclei give a nuclear-model-independent
bound on the $Z$-$e$-$\mu$ vertex which is twice
as strong as that obtained from $\mu\to e e e$.
In the case of E$_6$ models these limits provide  quite
stringent constraints on the $Z^\prime$ mass and on the $Z-Z^\prime$
mixing angle.
We point out that the proposed experiments to search for $\mu-e$
conversion in nuclei have good chances to find evidence of lepton flavour
violation, either in the case that
new exotic fermions are present at the electroweak scale, or
if a new neutral gauge boson $Z^\prime$ of E$_6$ origin
lighter than a few TeV exists.

\bs
\noindent
\vfill
\noindent
--------------------------------------------\phantom{-} \hb
\bigskip
\centerline{May 1993}

\bs\bs
\eject

\standardpage                            
\interlinea                              
\null
\noindent

\chap{Intro} 1. Introduction

The search for the conversion of muons into electrons in nuclei provides a
very stringent test of muon-number conservation. The present experimental
bound
on the branching for $\mu-e$ conversion in Titanium
$R\lesssim 4\times10^{-12}$ at TRIUMF [\cite{triumf}] and PSI [\cite{psi}],
gives a very powerful constraint on possible Flavour Changing Neutral Currents
(FCNCs) violating the muon and electron number conservation.
Due to the enhancement by the coherent
contribution of all the nucleons in the nucleus, the
limits on lepton flavour violation resulting from
this process are already more stringent than the ones
obtained from
the purely leptonic decays $\mu\to e e e$, $\mu\to e\gamma$, etc.
Furthermore, new experiments searching for $\mu-e$ conversion in nuclei are
planned, aiming to test
branching ratios up to $R\lesssim4\times10^{-14}$ [\cite{psi-new}], or
possibly even up to
$R\lesssim10^{-16}$ [\cite{melc}].
In the next few years, they could either provide the first accelerator
evidences
for lepton flavour violation, or give particularly strong constraints on
several possible extensions of the Standard Model (SM).

In the SM, Lepton Flavour Violating (LFV) currents are strictly
forbidden. This is not true in most of its extensions. For instance, if
right-handed neutrinos are present, LFV currents are generated radiatively,
proportional to very small GIM-like factors involving neutrino masses.
Other extensions of the SM which include new neutral
fermions and/or new Higgses, have been discussed in ref. [\cite{mueth}].
In model building, it is generally required that some natural mechanism
exists to suppress LFV currents at a level compatible with the present
experimental constraints.

Recently it has been stressed [\cite{enri}] that extended gauge models,
characterized by
additional $U(1)$ factors and by the presence of new charged fermions, predict
FCNCs mediated by the additional neutral gauge boson $Z'$.
Since the flavour changing $Z'$
vertices are expected to be naturally large,
these FCNCs must be
suppressed by a large $Z'$ mass.
In order to be consistent with
the limit on $\mu\to e e e$ and
for natural assumptions on the
fermion mass matrix the additional gauge boson
should not be much lighter than $\sim {\cal O}$(TeV)
[\cite{enri}].

In this paper we will consider the constraints implied by the present limit
on $\mu-e$ conversion in nuclei for the LFV currents mediated either by the
standard $Z$ boson, or by a new $Z'$.
By now these data have not
been used to constrain $Z'$ physics, and we show that
in most cases they give the strongest bounds on the FC
$Z'$ effects.
We also discuss the implications of the planned future
experiments [\cite{psi-new},\cite{melc}] on $\mu-e$ conversion in $Ti$.
If the underlying physics is described
by an extended gauge model like E$_6$, these experiments are expected
to reveal evidence for lepton flavour violation. If
no signal for LFV processes is detected,
this will result in very powerful constraints on
the structure of these models, implying vanishingly small
values for the
parameters describing fermion mixing, and/or
very large masses for the additional gauge bosons
($M_{Z'}\gtrsim 5$ TeV).
In section 2, we derive the effective LFV interaction between the charged
leptons and the nucleons, in terms of the fundamental lepton and quark neutral
current couplings. The $\mu-e$ conversion rate for the coherent nuclear
process is then obtained in a nuclear-model independent way.
Following ref. [\cite{enri}], in section 3 we show how possibly large FCNC
could naturally arise in extended gauge
theories. The case of E$_6$ models will be considered explicitly.
In section 4, we relate the E$_6$ parameters of section 3 to the effective
couplings relevant for the nuclear $\mu-e$ conversion process.
{}From the present experimental bound for $\mu-e$ conversion in $Ti$ we
derive new stringent constraints on the $Z$-$e$-$\mu$ vertex and on
the $Z'$ parameters, and we also discuss how these constraints will be
improved thanks to the proposed future experiments.
Finally in section 5 we present our conclusions.

\chap{Coher} 2. Coherent \bfmu{} -- \bfe{}  conversion

We will concentrate on the  case in which the LFV
interactions are mediated only by the exchange of massive gauge bosons,
and not by photon or scalar exchange.
In this case, the general lepton-quark effective
Lagrangian can be written in terms of a sum of contact interactions between the
leptonic and quark currents of the form
$$
{\cal L}_{eff}=
\sqrt{2}
G\, \bar e\gamma^\lambda(k_V-k_A\gamma_5)\mu
\sum_{q=u,d,s,...}\bar q\gamma_\lambda(v_q-a_q\gamma_5)q ,
\eq{effective}
$$
where $q=u,d,s,...$ are the relevant quark flavours.
$k_V,k_A$ are the LFV lepton couplings, and $v_q,a_q$ the
quark flavour diagonal couplings to the physical massive gauge boson
($Z$ or $Z'$) exchanged, which depend on the particular model considered.
For the contribution corresponding to
$Z$-exchange, $G=G_F$, the Fermi constant.
The $Z'$-exchange term has an overall strenght $G=G_FM_Z^2/ M_{Z'}^2$, and
whenever we will need to single out this case explicitly we will also
prime the couplings in eq. \req{effective}, $k_{V,A}\to k'_{V,A}$,
$v_q,a_q\to v'_q,a'_q$.

Since the maximum momentum transfer $q^2$ involved in the
$\mu-e$ conversion process is
much smaller than the scale associated with the structure of the
nucleon,  we can
neglect the $q^2$ dependence in the nucleon form factors.
Then, in the limit $q^2\approx 0$,
the matrix elements of the quark current for
the nucleon $N=p,n$       can be written as
$$
\eqalign{
<N\vert \bar q\gamma_\lambda q\vert N>=
G_V^{(q,N)} \bar N\gamma_\lambda N , \cr
<N\vert \bar q\gamma_\lambda\gamma_5 q\vert N>=
G_A^{(q,N)} \bar N\gamma_\lambda\gamma_5 N .\cr}
\eq{formfactors}
$$
In the limit in which strong isospin is a good symmetry,
that is
up to terms proportional to the up and down mass difference,
the neutron and proton form factors are related as follows
$$
\eqalign{
G^{(u,n)} = G^{(d,p)} \equiv G^{(d)} ,\cr
G^{(d,n)} = G^{(u,p)} \equiv G^{(u)} ,\cr
G^{(s,n)} = G^{(s,p)} \equiv G^{(s)} .\cr
}
$$

The conserved vector current and its coherent character, with the vector
charge equal to the quark-number, determine the couplings
$$
G_V^{(u)}= 2, \ \ \ \  G_V^{(d)}=1,  \ \ \ \   G_V^{(s)}=0.
\eq{2,3}
$$
This argument cannot be applied to the axial-vector current.
In terms of
definite $U(3)$-flavour transformation properties, one can introduce the
following combination of couplings
$$
\eqalign{
G_A^{(3)}=& G_A^{(u)}-G_A^{(d)}\cr
G_A^{(8)}=& G_A^{(u)}+G_A^{(d)}-2G_A^{(s)}\cr
G_A^{(0)}=& G_A^{(u)}+G_A^{(d)}+G_A^{(s)}.\cr
}
\eq{2,4}
$$
The weak currents transform as an octet under flavour $SU(3)$. The
two axial form factors
$G_A^{(3)}$ and $G_A^{(8)}$ can be expressed in terms of
the reduced amplitudes $F$ and $D$ extracted from the semi-leptonic decays
of baryons
$$
\eqalign{
G_A^{(3)}=& F+D=1.254\pm0.006\cr
G_A^{(8)}=& 3F-D=0.68\pm0.04.\cr}
\eq{2,5}
$$
The EMC [\cite{emc}] measurement of the polarization-dependent structure
function of the proton determines an additional independent combination of
$G_A^{(3)}$, $G_A^{(8)}$ and of the singlet $G_A^{(0)}$. One then obtains
$$
G_A^{0}=0.12\pm0.17
\eq{2,6}
$$
As a result all the axial form factors are determined.

At the nucleon level, the LFV Lagrangian \req{effective}
can then be written as
$$
{\cal L}_{eff}=
\sqrt{2} G\, \bar e\gamma^\lambda(k_V-k_A\gamma_5)\mu
\sum_{N=p,n}\bar N\gamma_\lambda(C_{1N}-C_{2N}\gamma_5)N ,
\eq{2,9}
$$
where the nucleon couplings are [\cite{pepe}]
$$
{\rm vector}: \qquad
\cases{
C_{1p} = 2 v_u + v_d ,\cr
C_{1n} =  v_u + 2v_d ,\cr }
\eq{2,10}
$$
and
$$
{\rm axial}: \qquad
\cases{
C_{2p} = G_A^{(u)}a_u+G_A^{(d)}a_d+G_A^{(s)}a_s ,\cr
C_{2n} = G_A^{(d)}a_u+G_A^{(u)}a_d+G_A^{(s)}a_s .\cr }
\eq{2,11}
$$
We will now discuss the
four nucleon couplings \req{2,10} and \req{2,11}
in the isospin formalism
for the nucleon, as appropriate for nuclear physics studies. Introducing
the nucleon spinor $\psi_N=\pmatrix{p\cr n\cr}$, and the isospin Pauli
matrix $\tau_3$, \req{2,9} reads
$$
\eqalign{
{\cal L}_{eff}=
&\sqrt{2} G\,
\bar e\gamma^\lambda(k_V-k_A\gamma_5)\mu      \cr
&\bar\psi_{N}\gamma_\lambda[(C_{1S}+C_{1V}\tau_3)-
(C_{2S}+C_{2V}\tau_3)\gamma_5]\psi_N,  \cr }
\eq{2,12}
$$
with the following couplings
$$
\eqalign{
{\rm Vector\ Isoscalar: }\qquad &
C_{1S} \equiv{1\over2} (C_{1p} + C_{1n}) = {3\over2}(v_u + v_d) \cr
{\rm Vector\ Isovector: }\qquad &
C_{1V} \equiv{1\over2} (C_{1p} - C_{1n}) = {1\over2}(v_u - v_d) \cr
{\rm Axial\ Isoscalar: }\qquad &
C_{2S} \equiv{1\over2} (C_{2p} + C_{2n}) = {1\over2}(G_A^{(u)}+G_A^{(d)})
                                           (a_u + a_d)+G_A^{(s)}a_s\cr
{\rm Axial\ Isovector: }\qquad &
C_{2V} \equiv{1\over2} (C_{2p} - C_{2n}) = {1\over2}(G_A^{(u)}-G_A^{(d)})
                                           (a_u - a_d).\cr
}\eq{2,13}
$$
At the low values of the squared momentum transfer
relevant for the kinematics of the $\mu-e$ conversion
process ($q^2\simeq -m_\mu^2$),
the matrix element of ${\cal L}_{eff}$ for a nuclear transition is
dominated by the coherent nuclear charge
associated with the vector current of the nucleon
$$
Q_W = (2Z+N)v_u + (Z+2N)v_d,
\eq{2,14}
$$
which gives an enhanced contribution to the coherent nuclear transition.
In  practice only the appropriate nuclear form factor for the coherent
contribution is needed.
The axial quark couplings $G_A^{(u,d,s)}$ do not contribute to the
coherent nuclear charge, and will only give rise to
nuclear-spin-dependent effects which are
negligible as long as the nucleon number $(A=Z+N)$ is large
enough.
For the nucleon numbers relevant for $\mu$-$e$ conversion experiments,
the rate for the coherent process, proportional
to $Q_W^2$, will indeed dominate over the incoherent
excitations of the nuclear system, which are sensitive to all the vector and
axial couplings given in Eq. \req{2,13}. This expectation is supported by
explicit calculations based on nuclear models
[\cite{kosmas-vergados-oset}], that  show that the
ratio between the coherent rate and the total $\mu-e$ conversion rate
for nuclei as $^{48}$Ti can be as large as 90\%.

In the non-relativistic limit for the motion of the muon in the muonic atom,
one can factorize the ``large" component of the muon wave function.
The corresponding coherent conversion rate is then given by
$$
\Gamma={G^2\over \pi}p_eE_e(k_V^2+k_A^2)Q_W^2\vert M(q)\vert^2,
\eq{coherate}
$$
where $p_e$ ($E_e$) is the electron momentum (energy), $E_e\simeq p_e\simeq
m_\mu$ for this process, and $M(q)$ is the
nuclear matrix element of the vector charge density
$$
M(q)=\int d^3x \rho(\vec x) e^{-i\vec q\cdot\vec x}\Phi_\mu(\vec x).
\eq{mq}
$$
In eq. \req{mq}, $\Phi_\mu(\vec x)$
is the normalized atomic wave function of the
muon and $\rho(\vec x)$ is the nuclear density (normalized to unity) taken
to be equal for proton and neutron distributions.

The form \req{coherate} is particularly convenient for discussing the
fundamental physics
involved in the $\mu-e$ conversion process, because it factorizes the
model dependent combination of couplings $(k_V^2+k_A^2)Q_W^2$ from the nuclear
matrix element squared. As said before, if both $Z$ and $Z'$ exchanges
mediate this FCNC process, then one has to reinterpretate the product
$(k_V^2+k_A^2)Q_W^2$, but not the nuclear ingredient factorization.

For nuclei with $A\lesssim100$ one can take, as customary in $\mu$-capture
analyses, an average value for the muon wave function inside the nucleus in
eq. \req{mq} in such a way that
$$
\vert M(q)\vert^2={\alpha^3 m_\mu^3\over\pi}{Z_{eff}^4\over Z}
\vert F(q)\vert^2,
\eq{mq2}
$$
where $Z_{eff}$ has been determined in the literature [\cite{zeff}]
and $F(q)$ is the
nuclear form factor, as measured for example from electron scattering
[\cite{escatt}].
One expects in $^{48}_{22} Ti$ this approximation to work
within a few percent, with $F(q^2\simeq -m_\mu^2)\simeq0.54$
and $Z_{eff}\simeq17.6$.

The branching ratio $R$ for $\mu-e$ conversion in nuclei normalized to the
total nuclear muon capture rate $\Gamma_{capture}$, which is
experimentally measured with a good precision, can then be computed
in any specific extension of the SM, and
the informations related to
the factors associated with new physics can be
extracted in a nuclear-model independent way.
In the case of FCNCs mediated by both $Z$ and $Z'$ exchange, we obtain
$$
R\simeq
{G_F^2\alpha^3\over\pi^2} m_\mu^3p_eE_e {Z_{eff}^4\over Z}
\vert F(q)\vert^2{1\over \Gamma_{capture}}\times
\eq{branching}
$$
$$
\phantom{R}
\left[(k_V^2+k_A^2)Q_W^2 +
2{M_Z^2\over M_{Z'}^2}(k_Vk'_V+k_Ak'_A) Q_W Q'_W +
\left(M_Z^2\over M_{Z'}^2\right)^2
({k'}^2_V+{k'}^2_A) {Q'}^2_W\right],
$$
where
$$
Q'_W=(2Z+N)v'_u + (Z+2N)v'_d
\eq{qpw}
$$
and we have explicitly primed the lepton and quark
couplings to the $Z'$ boson.
For $\Gamma_{capture}$ in $Ti$ we will use the
experimental determination
$\Gamma_{capture}\simeq (2.590\pm0.012)\times10^6 s^{-1}$
[\cite{mueexp}].

\chap{Fcnc} 3. FCNC in extended models

Following ref. [\cite{fit6}] we will now assume the effective low energy
gauge group is of the form
${\cal G} = [SU(2)_L \times U(1)_Y \times
SU(3)_C]\times U_1(1)$, and that it
originates  from the breaking of a {\it simple}
unification group, like E$_6$. The SM neutral gauge boson $Z_0$ can then
mix with the  $U_1(1)$ gauge boson $Z_1$, resulting in the two mass
eigenstates $Z$ and $Z'$. The NC Lagrangian in the physical $Z$ and $Z'$
basis  can be written as follows [\cite{fit6}],
$$
-{\cal L}_{\rm NC}=eJ^\lambda_{\rm em}A_\lambda +
g_0 (J^\lambda Z_{\lambda} + J'^\lambda Z'_{ \lambda}),
\eq{2.1}
$$
where $g_0 =(4\sqrt{2} G_F M^2_{Z_0})^{1/2}$ is the SM gauge coupling of
the $Z_0$, and $J$, $J'$ are the fermionic currents coupled to the $Z$ and $
Z'$ bosons. They are related to the gauge currents $J_0$ and $J_1$, coupled
to $Z_0$ and $Z_1$ respectively, by the rotation
$$
\pmatrix{ J^\lambda_Z \cr J^\lambda_{Z^\prime} } =
\pmatrix{\cos\phi&\sin\phi\cr -\sin\phi&\cos\phi\cr}
\pmatrix{ J^\lambda_0 \cr \sin\theta_w J^\lambda_1 },
\eq{2.5}
$$
where $\phi$ is the $Z-Z'$ mixing angle and $\theta_w$ is the weak mixing
angle.\note{We assume that the running of the $U_1(1)$ gauge coupling constant
$g_1$
from the unification scale down to low energy is similar to the running of the
hypercharge coupling constant. Normalising the $U_1(1)$
charge as the hypercharge generator $Y/2$ then yields
$g_1/g_0\simeq \sin \theta_w$.}

Besides predicting extra $Z'$ bosons,
extended gauge models like E$_6$ predict also
the existence of ``new" fermions $\psi_{\cal N}^0$.
The new fermions will in general mix with the standard
``known" fermions $\psi_{\cal K}^0$ having
the same electric and colour charges.
Then for any specific value
of the electric and colour charges,
the component of chirality $\alpha=L,R$ of the
light mass eigenstates $\psi_{l}$ will
correspond to a general superposition of
gauge eigenstates that can be written as  [\cite{fit6}]
$$
\psi_{l\alpha}=A^\dagger_\alpha\psi_{\cal K\alpha}^0 + F^\dagger_\alpha
\psi_{\cal N\alpha}^0.
\eq{2.3}
$$
The mixing matrices $A_\alpha$ and $F_\alpha$
describe respectively the mixing of the light states
with the known and the new fermions, and
satisfy the unitarity relation $A^\dagger_\alpha A_\alpha
+F^\dagger_\alpha F_\alpha=I$.
The presence of
these mixings will affect the couplings of the gauge bosons to the
light fermions $\psi_l$ [\cite{enri},\ccite{fit6}{ll-fit}].
In particular, given a general current $J_{\cal Q}^\lambda$,
corresponding to a broken generator ${\cal Q}$, its projection
on the light fermions $\psi_{l\alpha}$ will
read
$$
J^\lambda_{l{\cal Q}}
               =\sum_{\alpha=L,R}
\bar \Psi_{l\alpha} \gamma^\lambda \left[ q_\alpha^{\K} I +
(q_\alpha^{\N} - q_\alpha^{\K})
F^\dagger_\alpha F_\alpha \right]\Psi_{l\alpha}, \eq{2.10}
$$
\mn
where $q_{\alpha}^\K$ ($q_{\alpha}^\N$)
is the ${\cal Q}$-eigenvalue of the known (new) fermions
$\psi_{\K\alpha}^0$ ($\psi_{\N\alpha}^0$),
and for simplicity we have assumed that all the new states
have the same ${\cal Q}$-charge.
We refer to [\ccite{enri}{fit6}] for a more general discussion.

If the known fermions are mixed with new states
having different assignments of weak-isospin
(``exotic"  fermions), then the coefficient
$q^\N_\alpha-q_\alpha^\K=t_3(\psi_{\cal N\alpha}^0) -
t_3(\psi_{\cal K\alpha}^0)$
multiplying the mixing matrix $F^\dagger_\alpha F_\alpha$
in \req{2.10}  is non vanishing, and
the current $J_{l0}^\lambda$ coupled to the $Z_0$ boson is affected.
In this case extremely stringent bounds
on the off diagonal terms can be obtained from the limits on
FC processes. For example $(F^\dagger F)_{e\mu}\lesssim2\times
10^{-6}$ was obtained in [\cite{enri}]
from the non-observation of the $\mu\to ee e$ decay, however
we will see in section 4
that the limit from $\mu-e$ conversion in nuclei is stronger by a factor 2.
The diagonal elements of the matrix $F^\dagger F$ are also
constrained mainly from LEP, NC and charged current precision data
[\ccite{fit6}{ll-fit}] and the corresponding limits
are in general $\lesssim10^{-2}$.
On the other hand the mixings between the ordinary fermions and the
new exotic ones are theoretically expected to be very small, since
they arise in general from see-saw like formulas [\ccite{enri}{ll1}], so that
the corresponding limits are not very effective in constraining the
models under examination.

If instead the mixing is with new states having the same
$SU(2)$ assignments than the SM fermions
(``ordinary" fermions), the coefficient of the mixing term
in the $J_0^\lambda$ current is vanishing, and the couplings to the
$Z_0$ boson are not affected.
In this case no
phenomenological bounds can be set on the elements of $F^\dagger F$,
with the exception of the ordinary mixings of the
left-handed quarks, that are constrained by the
unitarity tests of the CKM matrix [\cite{ll-fit}].
However, ordinary-ordinary fermion mixing {\it does} affect the $J_1$
current, since in general $q_{1\alpha}^\N \neq q_{1\alpha}^\K $.
Clearly at low energy the possible effects of the ordinary-ordinary mixings
is suppressed with respect to the effects of the ordinary-exotic
mixings as the ratio of the gauge boson mass squared. However this
suppression could be largely compensated by the fact that in general these
mixings do not originate from see-saw like mass matrices, and then
{\it all} the entries in the
the mixing matrix $F^\dagger F$ can be large [\cite{enri}].


For definiteness we will now consider the case of E$_6$ models,
in which new gauge bosons as well as new ordinary and new exotic
fermions are present.
Since E$_6$ is rank 6, as many as two additional neutral gauge bosons could
appear in the low energy effective gauge group.
It is usefull to consider the embedding of the SM gauge group $\G$
in  E$_6$ through the following pattern of subgroups
$
\E \rightarrow \ U(1)_{\psi} \times SO(10)
\rightarrow \ U(1)_{\chi} \times SU(5)\rightarrow \G.
$
Then the lightest additional gauge boson
will in general correspond to an effective extra $U_1(1)$
resulting as a combination of the $U(1)_\psi$ and
$U(1)_\chi$ factors.
We will parametrize this combination in terms of an
angle $\beta$. This will define an entire
class of $Z^\pr$ models in which each fermion $f$ is coupled
to the new boson through the effective charge
$$
q_1(f)=q_\psi(f) \sin\beta + q_\chi(f) \cos\beta .
\eq{3.4}
$$
Particular cases that are commonly studied in the literature
[\cite{fit6},\cite{zp-new},\cite{rizzo-e6}]
correspond to $\sin\beta=-\sqrt{5/8}$, 0, 1
and are respectively denoted as $Z_\eta$, $Z_\chi$ and $Z_\psi$ models.
$Z_\psi$ occurs in $\E\to$ SO(10), while
$Z_\eta$ occurs in superstring models when
$\E$ directly breaks down to rank 5.
As we will see this model plays a peculiar role in the present analysis,
since it evades completely the kind of constraints that we are investigating.
Finally, a $Z_\chi$ boson occurs in SO(10)$\to$ SU(5) and
couples to the known fermions in the same way
than the $Z^\pr$ present in SO(10) GUTs.
However, since SO(10) does not contain additional charged fermions, the
kind of FC effects that we are studying here is absent.
In contrast, new charged quarks and leptons are present
in $\E$.  The
fundamental {\bf 27} representation contains,
beyond the standard 15 fermion degrees of freedom,
12 additional states for each generation,
among which we have a vector doublet of new leptons
$H=(N\> E^-)_L^T$, $H^c=(E^+ \> N^c)_L^T$.

The chiral couplings of the leptons to the $Z_1$ as well as the coefficient
of the LFV term $F^\dagger F$ are determined by the $q_\psi$ and $q_\chi$
charges of the new and known states, which are
$$
\eqlabel{3.8}
\eqalignno{
q_\psi(E_L)=-q_\psi(E_R)=-{1\o 3}\sqrt{5\o 2}, \hskip 1.2truecm
&q_\chi(E_L)=q_\chi(E_R)=-{1\o 3}\sqrt{3\o 2} ,  & \cr
q_\psi(e_L)=-q_\psi(e_R)={1\o 6}\sqrt{5\o 2}  ,\hskip 1.8truecm
&q_\chi(e_L)=3q_\chi(e_R)={1\o 2}\sqrt{3\o 2}.   & \req{3.8} \cr
}
$$
With respect to the $SU(2)_L$ transformation properties, the $E^+_L$ new
leptons are exotic and then the mixings of their CP conjugate states
$E^-_R$ with the standard R-handed leptons $e_R$ violates weak-isospin
by $1/2$. As is discussed for example in
Ref. [\cite{enri}] this kind of mixings
are generally suppressed as the ratio of the light to heavy
masses, and then for the $e$ and $\mu$ leptons
they are expected to be particularly small.
In contrast, the $E^-_L$ leptons are ordinary and their mixings
with the light leptons
are not expected to be suppressed by any small mass ratio since
they do not violate weak-isospin.
These mixings are generated by entries in the
mass matrix corresponding to v.e.v.s of singlet Higgs fields
$\langle \phi_S \rangle_0$ which,
since also contribute to the masses of the new (heavy) gauge
bosons,
are expected to be larger than the doublet v.e.v.s.
We note that in E$_6$ the ordinary-ordinary lepton
mixings occur between $SU(2)$ doublets, then it is clear that
 for each entry in the charged lepton
mass matrix of the form $E_R e_L \langle \phi_S \rangle_0$
there must be a corresponding entry $N^c \nu \langle \phi_S \rangle_0$
in the mass matrix for the neutral states, that would generate
a large Dirac mass for the light neutrinos.
Even if in the {\bf 27} of E$_6$ several new
neutral states (including two $SU(2)$ singlets) are present,
in the minimal E$_6$ models it is not possible
to generate naturally any small eigenvalue for the mass matrix
if these Dirac mass entries are present, since
the Higgs representation that could generate large Majorana masses and
lead to a see-saw mechanism is absent.
Then, in the frames of these models, the limits on the neutrino masses
automatically guarantee that any possible ordinary-ordinary mixing
in the charged lepton sector should be unobservably small.
However, as was discussed by Nandi and Sarkar [\cite{nandi-sarkar}],
large Majorana masses for the singlets  neutral
fermions can be generated due to gravitational effects,
leading to a rather complicated mass matrix for the neutral states
for which a see-saw mechanism is effective, and produce
naturally small masses for the light doublet neutrinos.
In this scenario, in order not to conflict with the limits on the
neutrino masses, there is no need to tune the Dirac mass entries to
any unnaturally small value. Then the weak-isospin conserving mixings
of the charged leptons are no more constrained, and in the limit in which
the singlet v.e.v.s are much larger than the doublet v.e.v.s are
theoretically expected to be ${\cal O}(1)$ [\cite{enri}].

The LFV lagrangian in E$_6$ models can be obtained from Eqs.
\req{2.1}, \req{2.10}.
For the charged leptons of the first two generations it reads
$$
- {\cal L}^{e\mu}_{LFV} =
g_0  [ k_0(\cos\phi Z_\lambda - \sin\phi Z'_\lambda)\bar
e_R\gamma^\lambda\mu_R +
k_1(\sin\phi Z_\lambda + \cos\phi Z'_\lambda)\bar e_L\gamma^\lambda\mu_L],
\eq{lfv}
$$ where
$$
k_0 = -{1\over2}(F_R^\dagger F_R)_{e\mu}
\eq{4.0a}
$$
is induced by the mixing with the exotic charged leptons $E^-_R$, while
$$
k_1=\sin\theta_w [q_1(E_L)-q_1(e_L)] (F_L^\dagger F_L)_{e\mu}
\eq{4.0}
$$
results from the mixing with the new ordinary leptons $E_L$.

{}From the second term in eq. \req{lfv},
we see that ordinary-ordinary fermion mixing can still
induce a LFV vertex
for the physical $Z$ boson. However this vertex is
suppressed by the $Z_0$-$Z_1$ mixing,
which is severely constrained by present data to
$\vert\phi\vert\lesssim0.02$ [\ccite{fit6}{zp-new}], and
then we can expect that in the presence of a ``light" $Z'$ the FCNC
processes would be mainly induced by direct $Z'$ exchange.


\chap{Constr} 4. Constraints from \bfmu{} -- \bfe{}  conversion in nuclei

The LFV parameters can now be constrained by
comparing the theoretical expression for the
branching ratio $R$ for the $\mu$-$e$ conversion process
in eq. \req{branching} to the experimental bound B.
Presently $B=4\times10^{-12}$ [\cite{triumf},\cite{psi}] at
90\% C.L., however we will also  discuss the limits
on the LFV parameters achievable with the planned future experiments.

First, the limits on $Z$-mediated FCNC can be obtained
in the limit in which the
$Z'$ is decoupled from low energy physics ($M_{Z'}\to\infty$ and
$\phi\to0$).
In this case, the quark vector couplings $v_f$ ($f=u,d$)
entering eq. \req{2,14} are given by the standard
expression $v_f=t_3(f_L)-2q_{em}(f) \sin^2\theta_w $.
We obtain
$$
(k_V^2+k_A^2) < 5.2\times10^{-13} \left(B\over 4\times10^{-12}\right),
\eq{boundzz}
$$
Independent limits on the LFV mixings of the R-handed
 or L-handed leptons can be given
in terms of the chirality
couplings $k_{L,R}={1\o 2}(k_V\pm k_A)$.
Then  \req{boundzz} implies
$\vert k_L^{e\mu}\vert,\vert k_R^{e\mu}\vert
<0.51\times10^{-6}$.
These limits are twice as strong as the corresponding
ones from the non observation of the decay $\mu\to e e e$
obtained in ref. [\cite{enri}].
In the case of
E$_6$ models the LFV couplings of the charged leptons to the $Z$ boson
originate only in the R-sector
($k_R=k_0$, $k_L=0$). From \req{boundzz} we obtain
$$
(F_R^\dagger F_R)_{e\mu}<1.0\times10^{-6}
\left( B\over4\times10^{-12}\right)^{1/2},
$$
that is tighter than the limit $(F_R^\dagger F_R)_{e\mu}<2.4\times10^{-6}$
from $\mu\to e e e$ [\cite{enri}].

As we see the limits from $\mu-e$ conversion in nuclei
on the LFV ordinary-exotic mixing of the first two
families are indeed quite strong.
We stress that due to the coherent enhancement of the rate, this process
gives the strongest constraint on the $Z-e-\mu$ vertex, twice more
stringent than that from $\mu\to e e e$.

However, as we have already discussed, these vertices are
expected to be suppressed as the ratio of the light and heavy
masses, that is by a factor of
the order ${m_\mu^2\over M_E^2}\lesssim 10^{-6}$ for $M_E\gtrsim100$
GeV. As a conclusion, at present these limits
are still not strong enough to effectively constrain
the models under
examination, since the possible FCNCs induced by such
naturally small ordinary-exotic mixings are still
compatible with the present  experimental data.

However the planned experiments [\ccite{psi-new}{melc}], aiming to
test branching ratios
down to $B\sim4\times 10^{-14}-10^{-16}$, do
have good chances to reveal signals of violation of lepton
flavour number induced by this kind of new physics.
If no signals are detected, the present limits will be improved to
$\vert k_L^{e\mu}\vert,\vert k_R^{e\mu}\vert <
0.51\times10^{-7}-0.25\times10^{-8}$
corresponding to a LFV ordinary-exotic mixing
$(F_R^\dagger F_R)_{e\mu}<(10-0.5)\times10^{-8}$.
This bound will indeed represent a serious constraint on E$_6$ models,
if the exotic states are assumed to be not much heavier than
the electroweak scale.

Let us now consider the effect of the mixing of the left-handed charged
leptons with the new ordinary states $E_L^-$ present in E$_6$.
In order to do this we will henceforth set the
ordinary-exotic mixing term
$(F_R^\dagger F_R)_{e\mu}$ to zero,
and we will concentrate on the consequences of having a non-vanishing
ordinary-ordinary mixing parameter
${\cal F}_{e\mu}\equiv(F_L^\dagger F_L)_{e\mu}$.
This is a safe procedure, since in the limit in which we neglect the
electron mass, there are no interference terms relating the L-handed
and R-handed lepton sectors, and the experimental limit on the
conversion of muons into electrons represents a fortiori a limit on
the production of electrons in the L-handed helicity state.

The LFV parameters
$k_V$ and $k_A$ entering eqs. \req{effective}-\req{branching},
can be read from eq. \req{lfv},
$$
\eqalign{
k_V=&k_A=k_1\sin\phi,\cr
k'_V=&k'_A=k_1\cos\phi,\cr
}
$$
while the quark couplings $v_f,v'_f$, $f=u,d$, entering in eqs.
\req{2,14}-\req{qpw}, are given by [\cite{fit6}]
$$
\eqalign{
v_f   =  &  \cos\phi[t_3(f_L)-2q_{em}(f)\sin^2\theta_w ] +
\sin\phi\sin\theta_w[q_1(f_L) + q_1(f_R)],
\cr
v'_f   = &   -\sin\phi[t_3(f_L)-2q_{em}(f)\sin^2\theta_w ] +
\cos\phi\sin\theta_w[q_1(f_L) + q_1(f_R)],
\cr}
\eq{21}
$$
where the $U_1(1)$ charge $q_1(f)$ that was
defined in \req{3.4} is given in terms of the
$q_\psi$ and $q_\chi$ charges for the quarks,
$$
\eqlabel{4.8}
\eqalign{
q_\psi(u_L)=-q_\psi(u_R)=q_\psi(d_L)=-q_\psi(d_R)={1\o 6}\sqrt{5\o 2},
                \cr
q_\chi(u_L)=-q_\chi(u_R)=q_\chi(d_L)={1\o 3}q_\chi(d_R)=
-{1\o 6}\sqrt{3\o 2}.    \cr
}
\eq{22}
$$
Due to the approximation made, for each value of the parameter
$\beta$ in \req{3.4}
the branching ratio \req{branching} depends only on the values of
$M_Z^\pr$, $\phi$ and ${\cal F}_{e\mu}\equiv(F_L^\dagger F_L)_{e\mu}$.
However, it is easy to see that since the gauge boson mixing effects in the
diagonal electron couplings are in any case very small
( $|\phi| \lesssim 0.02$ [\cite{fit6}-\cite{zp-new}]),
the relevant variables are actually only two, namely
${\cal F}_{e\mu} \cdot (M_Z^2/M_{Z^\pr}^2)$ and
${\cal F}_{e\mu} \cdot \phi$.
Moreover once the Higgs sector of the model
is specified, $M_{Z^\pr}$ and $\phi$ are
no more independent quantities.
For example an approximate relation that holds for
small mixings and when $M_{Z^\pr}$ ($\gg M_Z$)
originates from a large Higgs singlet v.e.v. [\cite{l-luo}] reads
$$
\phi \simeq - {M_Z^2\o M^2_{Z^\pr}} \sin\theta_w
{\sum_i t_3^i q_1^i |\langle\phi^i\rangle|^2
\o
\sum_i {t_3^i}^2 |\langle\phi^i\rangle|^2 },
\eq{3.9}
$$
and in this case the branching ratio \req{branching}
is in practice only a function of \break
${\cal F}_{e\mu}\cdot(M^2_Z/ M_{Z^\pr}^2)$.

The limits on the $Z'$ LFV parameter $M_{Z'}{\cal F}_{e\mu}^{-1/2}$,
obtained by comparing eqs.
\req{branching} to the present 90\% c.l. experimental bound
$B=4\times10^{-12}$ [\cite{triumf}], are plotted in Fig. 1. The thick
solid line depicts the limits
obtained by setting the gauge boson mixing angle $\phi$ to zero,
so that the
$\mu-e$ conversion is mediated only by $Z^\pr$ exchange in this case.
The resulting constraints are about twice as strong as the ones from
$\mu\to e e e$ found in ref. [\cite{enri}]. For most of the values of
$\sin\beta$, we find $M_{Z'}\left({{\cal
F}_{e\mu}\over10^{-2}}\right)^{-1/2}\gtrsim 5{\rm TeV}\times
\left(B\over4\times10^{-12}\right)^{-1/4}$.
Clearly it is not possible to translate the limits on the
$\mu$-$e$ conversion process directly into bounds on
$M_{Z'}$, since the value of the mixing parameter ${\cal F}_{e\mu}$ is
not known.
However, as we have discussed, from the theoretical point of view the
entries in the mixing matrix ${\cal F}$ are not expected to be
suppressed by any particularly small factor, and they are completely
unconstrained experimentally. Then it is reasonable to assume
$10^{-4}\lesssim{\cal F}_{e\mu}\lesssim10^{-1}$ as a natural range for
the ordinary-ordinary mixing parameter.
In this case, using the lower extreme ${\cal
F}_{e\mu}=10^{-4}$, we get a ``conservative naturalness" bound
$M_{Z'}\gtrsim 500$ GeV, for most of the values of $\sin\beta$. These
bounds are indeed quite strong, but since they are model dependent
obviously they cannot replace the
direct [\cite{zp-direct}] or indirect [\cite{fit6},\cite{zp-new}]
limits on the $Z^\pr$ parameters, which do not depend on any
assumption on the fermion mixings.

The planned experiment [\cite{melc}], aiming to
test the branching ratios  for the $\mu -e $ conversion process
down to $B\sim10^{-16}$, would allow
to improve the bounds up to
$M_{Z'}\gtrsim (5-100)$ TeV for the
same range of ``natural" values for ${\cal F}_{e\mu}$.
This would be a serious constraint on E$_6$ models,
and it is amusing to note that this kind of relatively unexpensive
experiments can in principle be sensitive to the presence
of a $Z'$ boson out of the reach of the supercolliders.


{}From Fig. 1 it is apparent that
two important exceptions are represented by the $\psi$
and the $\eta$ models,
corresponding respectively to $\sin \beta=-1$ and
$\sin \beta=-\sqrt{5/8}$, since in both these models the constraints
on the $Z'$ mass are evaded.

The absence of limits in the $\psi$ model is due to the fact that all
the standard fermions and their conjugate states belong to the same
representation of the $SO(10)$ subgroup of E$_6$, namely the {\bf 16},
and thus have the same $q_\psi$  abelian charge. As a consequence the
$q_\psi$ charges of the L- and R-handed states are equal and opposite
in sign, implying that the vector coupling to the $Z_\psi$ boson is
vanishing, and only the axial coupling is present. Then for this
particular value of $\beta$ it is not possible to obtain
strong bounds from $\mu-e$ conversion in nuclei.
In particular for $\phi=0$
no bounds at all are obtained on the parameter $M_{Z'}{\cal F}_{e\mu}^{-1/2}$
due to the fact that in the present analysis we have neglected the incoherent
contributions. In this case however a strong limit $M_{Z'}({\cal
F}_{e\mu} /10^{-2})^{-1/2}\gtrsim 3.7 $ TeV can still be obtained from the
non observation of the decay $\mu\to e e e$ [\cite{enri}].

The absence of limits in the $\eta$ model has quite a different
origin. Besides having $t_3^\K=t_3^\N$, the known and new ordinary
fermions also have $q_\eta^\K=q_\eta^\N$ for the particular value
$\sin \beta=-\sqrt{5/8}$. This implies that the coefficient of the
$F_L^\dagger F_L$ term is vanishing not only in the SM $J_0$ current,
but in the $J_1$ current as well. As a consequence any effect related
to the ordinary-ordinary mixing is completely absent in the $\eta$
model, independently of the kind of process considered. We refer to
ref. [\cite{enri}] for a more complete discussion on this point.

To study the possible effects on these results of a non vanishing
mixing angle $\phi$, i.e. when both the $Z^\pr$ and $Z$ bosons
contribute to the decay, we have used \req{3.9}
assuming, consistently with the conventional
E$_6$ models,  two doublets of Higgs fields with v.e.v.s $\bar v$  and $v$.
Since $\bar v$  and $v$ give mass respectively to the
$t$ and $b$ quarks, $\sigma \equiv {\bar v}^2/v^2 > 1$ is
theoretically preferred.
The bounds on $M_{Z^\pr}$ obtained by allowing
for a $Z_0$--$Z_1$ mixing consistent with this minimal Higgs
sector are shown in Fig. 1 by the dotted and dot-dashed lines,
which correspond to $\sigma = 1$ and $\infty$ respectively.
It is apparent that by allowing for a non vanishing
value of $\phi$, the limits on the $Z^\pr$ mass are qualitatively
unchanged.

Figure 2 depicts the constraints on the $Z'$ LFV parameter
$\phi\cdot{\cal F}_{e\mu}$.
The solid line shows the bounds obtained by
taking the limit $M_{Z^\pr} \rightarrow \infty$.
In this case the $\mu$-$e$ conversion process is mediated only by
the $Z$ boson, and is due to the mixing between the $Z_0$ and the $Z_1$.
It is apparent that the $Z_0$-$Z_1$ mixing angle is constrained
to be at most $\sim$ few$\,\cdot 10^{-4}/\left({\cal F}_{e\mu}\over 10^{-
2}\right) $ almost all over the $\beta$ axis.
For the smallest value of the mixing in the natural range
$10^{-4}\lesssim{\cal F}_{e\mu}\lesssim10^{-1}$, this is
comparable to the limit $\vert\phi\vert\lesssim10^{-2}$ resulting from the
fit to the available NC, charged current and LEP
data [\cite{fit6},\cite{zp-new}].
The dotted ($\sigma=1$) and dot--dashed lines ($\sigma=\infty$)
enclose the regions of the limits obtained assuming a minimal Higgs
sector.
In this case the value of $M_{Z^\pr}$ is finite and consistent,
according to \req{3.9}, with the values of $\phi$ at the bound.
We see that with this additional condition in practice the $Z$ and
$Z'$ bosons are constrained to be unmixed, except in a very small
region in the vicinity of the $\eta$ model.
The fact that in the case in which the Higgs sector is specified
the limits on $\phi$ are significantly tighter
than in the case in which
$\phi$ and $M_{Z^\pr}$ are assumed independent
(and the limit $M_{Z^\pr} \rightarrow \infty$ is taken)
means that the $\mu$-$e$ conversion in nuclei is
in first place sensitive to the $Z^\pr$ exchange, and thus
constrains the $Z^\pr$ mass,
while the contribution to the LFV transition of $Z_0$--$Z_1$ mixing alone
is less relevant and leads to loser constraints.
It is worth noting that this
behaviour is opposite to what is encountered in deriving
limits on the $Z^\pr$ parameters from precise electroweak data
[\cite{fit6},\cite{zp-new}],
where in fact the best bounds on the $Z^\pr$ mass are obtained
from the tight limits on $\phi$ implied by the LEP measurements.


\chap{Conclusions} 5. Conclusions

We have introduced the general charged lepton-quark contact Lagrangian
describing LFV neutral currents, we have derived the
corresponding effective lepton-nucleon interaction
and we have applied it to the case of $\mu$-$e$ conversion in muonic
atoms.
The relevant
nucleon vector couplings result from the coherent character of the
conserved vector current. The axial couplings are determined from
$SU(3)_f$ symmetry considerations and experiments, and their actual values
should be used to study the incoherent contribution to the
processes. However in the case of
$\mu-e$ conversion in nuclei with $A\gg 1$, the axial current
contribution can be neglected with respect to the
vector coherent contribution.
We have determined the rate of the coherent $\mu$-$e$ conversion process in
terms of the couplings appearing in the general lepton-quark effective
lagrangian, by means of the following additional approximations:
1) we have treated the muon as non-relativistic, which is correct up to
${\cal O}(\alpha Z)$; 2) we have taken an average for the $\mu$ wave function
inside the nucleus, which is a good approximation for $A\lesssim 100$;
3) we have used equal form factors for the proton and the neutron,
which is valid for light enough nuclei.
All these approximations work up to few percent for $^{48}_{22}Ti$.
We have then normalized the rate for $\mu-e$ conversion in nuclei with the
experimental value of the $\mu$-capture rate, rather than with the
theoretical expression which has beed previously
used in the literature [\cite{vergados}].

Following [\cite{enri}], we have discussed how
extended gauge models, predicting new neutral gauge bosons $Z'$
as well as new charged fermions, imply flavour changing
couplings between the $Z$ and $Z'$ gauge bosons and the known
fermions, and we have pointed out that in particular the $Z'$ flavour
changing vertices are expected to be unsuppressed.
As an example for illustrating this mechanism,
we have considered the case of E$_6$ models.

We have then studied the constraints on LFV couplings
from the limit on $\mu-e$
conversion in nuclei, obtaining the following results.

First, we have derived stringent bounds on the LFV interactions mediated
by the standard $Z$-boson, which in extensions of the SM can be
induced by the mixing
of the charged leptons with new exotic particles, and in particular in
E$_6$ models could appear in the right-handed leptonic sector.
The limits obtained are twice as strong as the ones from $\mu\to e e e$.
We have also discussed the sensitivity that will be attained by the
proposed future experiments searching for $\mu-e$ conversion in
nuclei, and we have shown that signals of
LFV transitions  induced by ordinary-exotic lepton mixing
are expected to be detected with these experiments
if the exotic leptons have masses not much larger than
the electroweak scale.

Second, we have considered the LFV interactions
induced in E$_6$ models by the mixing of the known charged leptons with new
ordinary states. In this case the
$\mu-e$ conversion proceeds through both $Z$ and $Z'$
exchange.
We have derived constraints on the relevant combinations of $Z'$
mass and mixing angle with the $Z'$ LFV couplings.
We have briefly discussed the reasons why the $Z'$ LFV couplings are
theoretically expected to be large, and we have concluded that
in order to account for the
non observation of $\mu-e$ conversion in nuclei, the $Z'$
should be sufficiently heavy (in most cases at least at the TeV scale)
to suppress the transition rate, and almost unmixed with the standard $Z$.

We have suggested that
the simultaneous presence of new charged fermions and
new gauge bosons with mass up to a few TeV should give rise
to LFV transitions that should be observed in future
experiments looking for $\mu-e$ conversion in nuclei
with improved sensitivity.
On the other hand, if no effect were
found, the resulting limits on these kind of FCNCs will be extremely
severe, implying in most cases $M_{Z'}\gtrsim5$ TeV
unless the LFV couplings are tuned to be smaller
than $\simeq 10^{-4}$.

As we have discussed in some detail,
the constraints on the $Z'$ mass
presented here do not apply to two particular
E$_6$ models.
In the $\psi$ model the quark vector couplings to the $Z'$ vanish, so that
there is no coherent contribution to $\mu-e$ conversion in nuclei, and
then leptonic processes like $\mu\to e e e$ should be used
to constrain the possibly large $Z'$-mediated FCNCs.
On the other hand, as was already stressed in [\cite{enri}],
in the superstring-inspired $\eta$ model
the large $Z'$-mediated LFV are completely absent,
implying that the kind of
constraints discussed here are not effective to derive limits on the
$Z_\eta$ parameters independently of the
particular experimental process considered.

\chap{Ack} Acknowledgements

We would like to thank Prof. V.M. Lobashev for stimulationg discussions on
this topic. One of us (D.T.) acknowledges a post-doc fellowship from the
Spanish Ministry of Education and Science. This work has been partially
supported by CICYT (Spain), under grant AEN/90-0040.


\def\ea{{\it et al.}}
\def\ib{{\it ibid.\ }}

\def\np#1{Nucl. Phys. {\bf #1}}
\def\npa#1{Nucl. Phys. {\bf A#1}}
\def\npb#1{Nucl. Phys. {\bf B#1}}
\def\plb#1{Phys. Lett. {\bf B#1}}
\def\pr#1{Phys. Rev. {\bf #1}}
\def\prd#1{Phys. Rev. {\bf D#1}}
\def\prc#1{Phys. Rev. {\bf C#1}}
\def\prl#1{Phys. Rev. Lett. {\bf #1}}
\def\zpc#1{Z. Phys. {\bf C#1}}
\def\prep#1{Phys. Rep. {\bf #1}}

\null
\baselineskip 8pt
\centerline{\title References}
\vskip .8truecm

\biblitem{triumf}
D.A. Bryman \ea, \prl{55} (1985) 55;\hbup
S. Ahmad \ea, \prl{59} (1987) 970;\hbup
S. Ahmad \ea, \prd{38} (1988) 2102.\par

\biblitem{psi}
A. Badertscher \ea, J. Phys. {\bf G17} (1991) S47;\hbup
A. Van der Schaaf, Nucl. Phys. {\bf A546} (1992) 421C.\par

\biblitem{escatt}
I. Sick and J.S. McCarthy, \npa{150} (1970) 631; \hbup
W. Bertozzi, J. Friar, J. Heisenberg and J.W. Negele, \plb{41} (1972)
408;\hbup
W. Bertozzi \ea,  \prl{28} (1972) 1711;\hbup
B. Dreher \ea, \npa{235} (1974) 219;\hbup
W. Donnely and J.D. Walecka, Ann. Rev. Nucl. Sci. {\bf 25} (1975) 329;\hbup
B. Frois and C.N. Papanicolas,
Ann. Rev. Nucl. Sci. {\bf 37} (1987) 133.\par

\biblitem{psi-new}
A. Badertscher \ea, SINDRUM II collaboration, preprint PSI-PR-90-41
(1990).\par

\biblitem{melc}
V.S. Abadjev \ea, MELC collaboration, preprint MOSCOW (1992).\par

\biblitem{pepe}
J. Bernab\'eu, Z. Phys. {\bf C56} (1992) S24.\par

\biblitem{mueexp}
T. Suzuki, D.F. Measday and J.P. Roalsvig, \prc{35} (1987) 2212.\par

\biblitem{mueth}
W.J. Marciano and A.I. Sanda, \prl{38} (1977) 1512;\hbup
J.D. Vergados,  \prep{133} (1986) 1;\hbup
J. Bernab\'eu, An. F\'\i s. {\bf 5} (1989) 36;\hbup
T.S. Kosmas and J.D. Vergados, Nucl. Phys. {\bf A510} (1990) 641.\par

\biblitem{enri}
E. Nardi, preprint UM-TH 92-19 (1992), to appear in \prd{}.\par

\biblitem{emc}
J. Ashman \ea, EMC experiment, \plb{206} (1988) 364;
\npb{328} (1989) 1.\par

\biblitem{kosmas-vergados-oset}
T.S. Kosmas and J.D. Vergados in ref. [\cite{mueth}];\hbup
H.C. Chiang, E. Oset, T.S. Kosmas, A. Faessler and J.D. Vergados,
submitted to \npa{}.\par

\biblitem{zeff}
J.C. Sens, \pr{113} (1959) 679;\hbup
K.W. Ford and J.G. Wills, \np{35} (1962) 295;\hbup
R. Pla and J. Bernab\'eu, An. F\'is. {\bf 67} (1971) 455;\hbup
H.C. Chiang \ea, in Ref. [\cite{kosmas-vergados-oset}].

\biblitem{ll-fit}
P. Langacker and D. London, \prd{38} (1988) 886.\hbup
E. Nardi, E. Roulet and D. Tommasini,  \npb{386} (1992) 239; \hbup
E. Nardi, E. Roulet and D. Tommasini,
in ``Electroweak Physics beyond the Standard
Model", ed J.W.F. Valle and J. Velasco, World Scientific (1992).\par

\biblitem{nandi-sarkar}
S. Nandi and U. Sarkar, \prl{56} (1986) 564;\hbup
M. Cveti\u c and P. Langacker, \prd{46} (1992) R2759.\par

\biblitem{vergados}
See e.g. J.D. Vergados in ref. [\cite{mueth}].\par

\biblitem{zp-fc}
T.K. Kuo and N. Nakagawa, \prd{32} 306 (1985).\par

\biblitem{e6-fcloop}
J. Bernab\'eu \ea, \plb{187} 303 (1987). \par

\biblitem{e6-fcrizzo}
G.Eilam and T.G. Rizzo, \plb{188} 91 (1987). \par

\biblitem{salati-trip}
B.W. Lee, \prd{6} 1188 (1972); \hbup
J. Prentki and B. Zumino, \npb{47} 99 (1972); \hbup
P. Salati, \plb{253} 173 (1991). \par

\biblitem{zp-direct}
CDF Collaboration, F. Abe \ea, \prl{67} (1991) 2609; \ib {\bf 68},
(1992) 1463. \par

\biblitem{fit6}
E. Nardi, E. Roulet and D. Tommasini,
\prd{46} (1992) 3040. \par

\biblitem{zp-new}
J. Layssac, F.M. Renard and C. Verzegnassi, \zpc{53} (1992) 97; \hbup
M.C. Gonzalez Garc\'\i a and J.W.F. Valle; \plb{259} (1991) 365; \hbup
G. Altarelli et al., \plb{263} (1991) 459; \hbup
F.del Aguila, J.M. Moreno and M. Quir\'os, \npb{361} (1991) 45; \hbup
F. del Aguila, W. Hollik, J.M. Moreno and M. Quir\'os, \ib {\bf B372}
 (1992) 3;\hbup
P. Langacker and M. Luo, \prd{45} (1992) 278. \par

\biblitem{l-luo}
P. Langacker and M. Luo, in Ref. [\cite{zp-new}].\par

\biblitem{rizzo-direct}
P. Langacker and M. Luo, in Ref. [\cite{zp-new}];  \hbup
T.G. Rizzo; talk given at the 15$^{th}$ Johns Hopkins
Workshop on Current Problems in Particle Theory, Baltimore, MD, August
26-28 1991; ANL-HEP-CP-91-85. \par

\biblitem{fit}
E. Nardi, E. Roulet and D. Tommasini,  \npb{386} (1992); \hbup
E. Nardi, E. Roulet and D. Tommasini,
in ``Electroweak Physics beyond the Standard
Model", ed J.W.F. Valle and J. Velasco, World Scientific (1992).\par

\biblitem{ll1}
P. Langacker and D. London, in ref. [\cite{ll-fit}].\par

\biblitem{ll2}
P. Langacker and D. London, \prd{38} (1988) 907.\par

\biblitem{slansky}
For a review on the group E$_6$, see R. Slansky; \prep{79} (1981) 1. \par

\biblitem{pdg92}
Particle Data Group, J.J. Hern\'andez {\it et al.},
\prd {\bf 45} Part. II (1992). \par

\biblitem{sindrum}
SINDRUM collaboration, U. Bellgardt \ea, \npb{299}
(1988) 1. \par

\biblitem{rizzo-e6}
J.L. Hewett and T.G. Rizzo, \prep{183} (1989) 195, and references
therein. \par

\biblitem{rosner86}
D. London and J.L. Rosner, \prd{34} (1986) 1530. \par

\interlinea


\insertbibliografia

\par\vfill\eject

\centerline{\lltitle Figure captions}
\bigskip
\baselineskip 14pt

\medskip\noindent
Fig. 1:
Limits on the $Z^\prime $ LFV parameter
$M_{Z^\prime}\cdot {\cal F}_{e\mu}^{-1/2}$
from the experimental limits on the $\mu$-$e$ conversion process,
for a general $\E$ neutral gauge boson, as a
function of $\sin\beta$.
The mixing term ${\cal F}_{e\mu}$ is given in units
of $10^{-2}$. The vertical units are TeV when
the current limit on the branching for $\mu$-$e$ conversion
$B=4\cdot 10^{-12}$ is taken.
The limits on the $Z^\prime$ mass for different values of
the experimental branching ratio and/or of
${\cal F}_{e\mu}$ can be easily read off the figure by
properly rescaling the vertical units.
The thick solid line is obtained by setting
the $Z_0$--$Z_1$ mixing angle $\phi$ to zero.
The bounds obtained by
allowing for a non-vanishing $Z_0$--$Z_1$ mixing, consistent
with the values of $M_Z^\prime$ when a minimal Higgs sector is assumed, are
also
shown. The dotted lines correspond to equal vevs of the two Higgs
doublets present in the model, {\it i.e.}
$\sigma\equiv {\bar v}/v=1$ while the dot--dashed lines
correspond to $\sigma=\infty$.

\bs\bs\bs
\medskip\noindent
Fig. 2:
Limits on the $Z^\prime$ LFV parameter
$\phi \cdot {\cal F}_{e\mu}$
from the experimental limits on the $\mu$-$e$ conversion process,
for a general $\E$ neutral gauge boson, as a
function of $\sin\beta$.
The current limit on the branching for $\mu$-$e$ conversion
$B=4\cdot 10^{-12}$ is assumed and
the mixing term ${\cal F}_{e\mu}$ is given in units
of $10^{-2}$.
The limits on the $Z_0$--$Z_1$ mixing angle $\phi$
for different values of
the experimental branching ratio and/or of
${\cal F}_{e\mu}$ can be easily read off the figure by
properly rescaling the vertical units.
The thick solid lines are obtained in the
limit $M_{Z^\prime}\rightarrow \infty$. The dotted $(\sigma=1)$ and dot-dashed
$(\sigma=\infty)$ lines show the limits obtained for a finite
$Z^\prime$ mass and assuming a minimal Higgs sector.

\vfill\eject
\bye